\documentclass{article}
 		
\newcommand{\msh}{MSH~11-61A }
\newcommand{\g}{$\gamma$}
\newcommand{\pio}{$\pi^o$\ }

\newcommand{\erg}{\ {\rm erg}}
\newcommand{\eV}{\ {\rm eV}}
\newcommand{\MeV}{\ {\rm MeV}}

\newcommand{\TeV}{\ {\rm TeV}}
\newcommand{\kpc}{\ {\rm kpc}}
\newcommand{\pc}{\ {\rm pc}}
\newcommand{\ve}{\varepsilon}
\newcommand{\veg}{\varepsilon_\gamma}
\renewcommand{\AA}{{\it A\&A}\ }

\newcommand{\AAS}{{\it \AA Suppl.}\ }
\newcommand{\ApJ}{{\it ApJ}\ }
\newcommand{\ApJS}{{\it \ApJ Suppl.}\ }

\newcommand{\Nature}{{\it Nature}\ }
\newcommand{\MNRAS}{{\it MNRAS}\ }
\newcommand{\PASJ}{{\it PASJ}\ }
\newcommand{\PASP}{{\it PASP}\ }

\begin{document}

\ \vskip1.4cm

\leftline{\bf A NEW MODEL FOR THE THERMAL X-RAY COMPOSITES}
\leftline{\bf AND THE PROTON ORIGIN \g-RAYS FROM SUPERNOVA} 
\leftline{\bf REMNANTS}

\vspace{1cm}
\noindent
\hspace{2cm}O.~PETRUK\par\noindent
\hspace{2cm}{\it Institute for Applied Problems in Mechanics and Mathematics}
\par\noindent
\hspace{2cm}{\it 3-b Naukova St., 79053 Lviv, Ukraine}
\vspace{1.4cm}

\hspace{-0.67cm}
{\small 
\begin{tabular}{@{}p{11.7cm}}
{\bf Abstract}\ 
Recent nonthermal X-ray and \g-ray
observations, attributed to electron emission processes, for the first 
time give an experimental confirmation that electrons are accelerated on SNR 
shocks up to the energy $\sim 10^{14}\eV$. 
We have no direct observational confirmations about 
proton acceleration by SNR. Different models of \g-emission from SNRs predict 
different emission mechanisms as dominating. Only \pio decays created in
proton-nucleon interactions allow us to look inside the CR nuclear component 
acceleration processes. A new model for the thermal X-ray composites 
strongly suggest that thermal X-ray peak inside the radio shell of SNR 
tells us about entering of one part of SNR shock into a denser medium 
compared with other parts of
the shell. This makes a TXCs promising sites for \g-ray generation 
via \pio decays. Detailed consideration of SNR-cloud interaction allows to increase
an expected proton induced \g-ray flux from SNR at least on an order of magnitude,
that allows to adjust the theoretical \pio decay \g-luminosities 
with observed fluxes at least for a few SNRs even for low density 
($n_o\sim 10^{1}\div10^{2}\ {\rm cm^{-3}}$) cloud. 
\end{tabular}
}
\vspace{0.8cm}

\section{Introduction}

Since SNRs possess high enough energy, their shocks are belived to be a major 
contributor of both electron and nuclear components of Galactic CRs 
up to energies $10^{15}\eV$ \cite{CRbooks}.  
SNRs emit in all wavelengths. Their radio emission gives clear evidence about 
relativistic electrons 
accelerated on the shock front. 
Observed nonthermal X-ray emission from several SNRs 
(SN 1006 \cite{Koyama_et_al-95}, Cas~A  \cite{Allen_et_al-97}, Tycho 
\cite{Ammosov-et-al-94}, 
G347.3-0.5 \cite{Koyama_et_al-97}, IC443 \cite{Keohane-P-97}, 
G266.2-1.2 \cite{Slane-Huges-et-al-00}) 
is thought to be also synchrotron, from the shock accelereted electrons 
with a much higher energy ($\sim 10^{13}\eV$). 
Observed synchrotron 
X-ray \cite{Koyama_et_al-95} and TeV \g-ray \cite{Tanimori_et_al-98} 
emission from 
SN~1006 as well as from G347.3-0.5 
\cite{Slane-et-al-99,Muraishi-et-al-00}, 
by attributing the emission to the synchrotron radiation, 
gives firm confirmations that electrons are accelerated on the SNR shocks 
up to energies $\sim 10^{14}\eV$. 
The theory of acceleration of CRs on the shock front via 
the first order Fermi mechanism \cite{
CRbooks}  
predicts that the number of electrons $N_e$ involved in the acceleration 
process is much smaller than the number of protons $N_p$: 
$N_e(\ve)/N_p(\ve)=(m_e/m_p)^{(\alpha-1)/2}\simeq 10^{-2}$ ($\alpha$ is the spectral index of
accelerated particles, $\ve$ is the energy of particles).   
Thus, if there are electrons accelereted to such high energies, 
we should expect that protons with the same high energy have to reveal 
themselves in observations, too. 
We have to look for the objects which might emit \g-rays mainly via 
neutral pion decays.    

\section{Observations of the \g-rays from SNRs}

The visibility of SNRs in Gev and TeV \g-rays was treated in 
\cite{Drury-Ah-V-94,Ahar-Dr-V-94}. The main conclusion from the theory 
is: if SNRs are mainly responsible for the galactic CRs, it 
may be difficult to observe SNRs by EGRET, but 
should be possible to detect compact (with angular size $\sim 0.25'$) 
remnants from a distance less than several kpc in 
TeV band, with Cherenkov telescopes.  
Unfortunately, as first results on TeV \g-observations 
\cite{Hillas-95} as 
well as most of the next (G78.2+2.1, W28, IC~443, \g-Cygni, W44, W63, Tycho; 
see short reveiw in \cite{Rowell-et-al-00}) 
have given a negative result. 
Till now TeV \g-rays were observed from only two SNRs.   

SN~1006 is the first shell-like SNR detected as a $\TeV$ source 
\cite{Tanimori_et_al-98}. 
Whereas authors in \cite{Tanimori_et_al-98} attributed this emission to 
synchrotron, the emission is
tried to be explained by other mechanisms, too.  
Recent discussion on the origin of $\TeV$ \g-rays form SN~1006 
\cite{Ahar-Ato-99} shows that a proton origin of \g-rays is
possible if distance to the remnant is about $1\kpc$ 
and there is a significant compression of gas. 
Nevertheless other
emission mechanisms may not be excluded (see, e. g.
\cite{Reynolds-96}). The observations of G347.3-0.5 reported recently  
\cite{Muraishi-et-al-00} also reveal a TeV \g-flux. Authors 
attributed this emission to inverse Compton scattering of CMBR by 
shock accelerated electrons and estimate \pio induced contribution as too low, 
but did not exclude it because of possibility of interactions with a 
cloud \cite{Slane-et-al-99}. 

The first \cite{1EG}, second \cite{2EG} and the third \cite{3EG} EGRET 
catalogue of the high-energy \g-ray sources 
among
the general categories (solar flare, pulsars, \g-ray bursts, radio galaxy, 
active galactic nuclei and blazars), include lists of sources for which no 
identification with objects at other wavelengths is yet surely found. 
The analysis of possible associations of  
the SNR positions 
with error circles of the unidentified \g-sources was first performed in
\cite{Sturner-Dermer-95}. 
Analysis of unidentified sources from the third EGRET catalog 
\cite{Romero-B-T-99} 
shows 22 such possible assosiations with probability of being of chance less than
$10^{-5}$. 
Some other unidentified EGRET 
sources might be related to yet undetected SNR \cite{Combi-R-B-98}.
Of course, a part of these remnants have or may have their compact stellar 
remnants \cite{Chevalier-93}, 
pulsars, which can be responsible for the \g-ray emission. 
There are 225 known SNRs in our Galaxy \cite{Green-2000}. 
The short history of studies on association of EGRET sources with SNRs 
may be found in \cite{Romero-B-T-99}.  

Up to now, there is no clear observational confirmation that 
nuclear component of CRs is generated by SNRs.   
There is only one observation 
reported 
as confirmation of this thought \cite{Combi-R-B-98}. 
The nature of \g-emission in all observations 
is very questionable. 

There are several emission processes in competition in analysis of observed 
SNR \g-ray spectra. It depends on the conditions 
in emission site and on the way to the observer which one will dominate.     
It is well known that we should have 
a high number density of target nuclei 
in order to make the \pio decay mechanism dominate in a model. 
Therefore, \g-rays from proton-proton collisions are expected from the SNRs 
which are located near the dense interstellar material and reveal evidence 
about interaction with it. 

\section{Thermal X-ray composites. A new model in 3 dimensions}

To look for signatures of proton acceleration in SNR, it is interesting 
to consider a mixed-morphology class of SNRs which is known also as thermal 
X-ray composites (TXCs). These are remnants with a thermal X-ray centrally filled 
morphology within the radio-brightened limb. Such remnants were first 
reviewed in \cite{Seward-85}. The authors of \cite{Rho-Petre-98} 
argued that they create a 
separate morphological class of SNR. It is interesting that most of these SNRs  
reveal observational evidence about the interaction with nearby 
molecular clouds.  
There are 7 remnants in the analysis of assosiation of the 2EG sources with SNRs 
\cite{Sturner-Dermer-M96}. Four of them (W28, W44%
, MSH~11-61A, IC443) belong to the class mentioned and 
interact with clouds. 

The authors of \cite{Rho-Petre-98} have emphasized two prominent 
morphological distinctions of TXCs: 
a) the X-ray emission is thermal; the 
distribution of X-ray surface brightness 
is centrally peaked and fills the area within the radio shell, 
b) the emission arises primarily from the swept-up ISM material, not from 
ejecta. 
Besides similar morphology, the sample 
of SNRs also has similar physical properties \cite{Rho-Petre-98,Pet-2000b}. 
Na\-me\-ly, 
i) the same or higher central density compared to the edge, 
ii) complex interior optical nebulosity,  
iii) higher emission measure 
in the region of X-ray peak localisation, 
v) temperature profiles are close to uniform. 
Seven objects from the list of 11 TXCs interact with molecular clouds 
\cite{Rho-Petre-98}. 
Thus, ambient media in the regions of their location are nonuniform 
and cause a nonsphericity of SNRs. 

Two physical models have been presented to explain TXC (see review in 
\cite{Rho-Petre-98,Pet-2000b}). 
These models are used to obtain the centrally 
filled X-ray morphology within the 
framework of one-di\-men\-si\-o\-nal (1-D) hydrodynamic approaches. 
When we proceed to 2-D or 3-D hydrodynamical 
models, we note that a simple projection 
effect may cause the shell-like SNR to fall into another morphology class, 
namely, to become TXC \cite{Hn-Pet-99}. 
The main feature of such a SNR is the thermal X-rays 
emitted from the swept-up gas and 
peaked in the internal part of the projection. 
Densities over the surface of a nonspherical SNR may essentially 
differ in various regions. 
If the ambient density distribution 
provides a high density in one of the 
regions across the SNR shell and is high enough to 
exceed the internal column density near the edge of the projection, 
we will see a centrally filled X-ray projection. 
Thus, this new model strongly suggests that the thermal X-ray peak inside the 
radio shell
tells us that one part of SNR shock has entered into 
a denser medium compared with other 
parts. This makes TXCs promising sites for \g-ray generation 
via \pio decays.  

\section{SNR \g-rays from decay of neutral $\pi$-mesons}

Whereas astrophysical realisations of several emission mechanisms allow us to
make conclusions about acceleration of CR electron component on the
shock fronts of SNRs, only \pio decay \g-rays \cite{CRbooks} 
give us the possibility to look at the 
proton component acceleration processes having data in $\veg\geq 100\MeV$.   

\subsection{Estimations on the $\pi^o$ decay \g-ray luminosity}

Luminosity of an SNR in $\pi^o$ decay \g-rays 
in $\varepsilon\geq\varepsilon_{\min}/6$ band  
is \cite{Sturner-Dermer-95,Hn-Pet-98}: 
\begin{equation}
L_\gamma={c\overline{\sigma}_{pp}\over 6}n_N W_{\rm cr}.
\label{Lumin}
\end{equation}
where $\varepsilon_{{\min}}\approx600$ MeV is the minimal proton kinetic 
energy of the effective pion creation, 
cross section $\sigma_{pp}(\varepsilon)$ is 
close to the mean value $\overline{\sigma}_{pp}=3\cdot10^{-26}$ cm$^2$, 
$n_N$ is the mean number density of target nuclei 
and $W_{\rm cr}$ is the total energy of cosmic rays in the SNR with 
$\varepsilon\geq\varepsilon_{\min}$.
There are different estimations of the efficiency $\nu$ of the flow's 
kinetic energy transformation into the energy of accelerated particles: 
$W_{\rm cr}=\nu E_o$. 
We take acceptable value $\nu=0.03$%
.
Thus, in the first approach, the theoretical estimation on the 
$\pi^o$ decay \g-luminosity of any SNR is 
$$
L_{\gamma}^{\geq 100}=6.3\cdot10^{33}\overline{n}_o\nu_3E_{51}\quad {\rm erg/s},
$$
where $\nu_3=\nu/0.03$, 
$E_{51}$ is the energy of supernova explosion $E_o$ in the units of  
$10^{51}\ {\rm erg}$, 
$\overline{n}_o=\overline{n}^o_N/1.4$ is the average hydrogen number 
density within SNR which 
equals to average hydrogen number density of the ambient medium 
in the region of an SNR location, in ${\rm cm^{-3}}$. 

The real situation is more complicated. Often 
only a part of SNR interacts with a denser ISM 
material. There are factors 
which increase CR energy density $\omega_{cr}$ 
\cite{Hn-Pet-98}: 
1) The CRs are not uniformly distributed inside an SNR; 
most of CRs are expected to be in a thin shell 
near the shock front where most of swept-up mass is 
concentrated. 
2) The reverse shock from interaction with dense cloud 
also increases the energy density of CRs. 
These factors enhance $\omega_{cr}$ in the region of 
interaction \cite{Hn-Pet-98}:
\begin{equation}
\omega_{\rm cr}\approx1.7\left(\gamma\over\gamma-1\right)^{3/2}
\overline{\omega}_{\rm cr}\ ,
\label{omega}
\end{equation}
where CR energy density in the region of interaction is 
$\omega_{\rm cr}=W_{\rm cr}/V_{\rm int}$ and 
$\overline{\omega}_{\rm cr}=W_{\rm cr}/V_{\rm snr}$. This gives  
$\omega_{\rm cr}\approx 6.6\ \overline{\omega}_{\rm cr}$ for $\gamma=5/3$.
 
If we put $W_{\rm cr}=\omega_{\rm cr}V_{\rm int}$
into (\ref{Lumin}), we obtain  
with (\ref{omega}) that for any SNR  
\begin{equation}
L_\gamma^{\geq 100}=1.7\cdot10^{35}\eta n_o\nu_3E_{51}
\quad {\rm erg/s}
\label{lum-eta}
\end{equation}
where $\eta=V_{\rm int}/V_{\rm snr}$, $n_o$ is the number 
density of the ambient medium before the shock wave 
in the region of interaction, 
$\gamma=5/3$. We have to
take into account that region of interaction is not extended to the region 
before 
the shock, since the energy density of CR should be considerably lower outside 
the SNR \cite{CRbooks,Hn-Pet-98}. 
It is easy to estimate $\eta$ following the consideration in \cite{Hn-Pet-98}: 
\begin{equation}
\eta=0.18\mu^2\sqrt{\xi},\qquad \mu\leq0.2,
\label{eta}
\end{equation}
where $\xi=n_o/\overline{n}_o$, $\mu=R_{int}/\overline{R}$ , 
$R_{int}$ is the avarage radius of the surface 
of interaction, $\overline{R}$ is the avarage radius of SNR. 

Thus, considering the hydrodynamic process of SNR-cloud interaction in details, 
it is possible to increase the expected \pio decay \g-ray flux 
by $26\eta\xi\simeq0.2\xi^{3/2}$ times, i. e., up to few orders of magnitude. 
This is
enough e.g. for explanation of TeV \g-rays from SN1006 as a result of decays of
neutral pions \cite{Tanimori_et_al-98}. 
Additional factor increasing the effectivity of \pio meson production 
is instability of 
contact discontinuity between the SNR and the cloud material which 
mixes the media with high energy particles and target nuclei \cite{Hn-Pet-98}. 
This factor should also be considered in future.

\subsection{Nucleonic origin \g-rays from MSH~11-61A}

SNR G290.1-0.8 (MSH~11-61A) is located in the southern hemisphere. 
The distance to the remnant, $7\kpc$, is obtained from the optical 
observations \cite{Rosado-Ambr-96}, but not yet confirmed by X-ray observations 
\cite{Slane-et-al-msh}. X-ray and radio morphologies 
\cite{Seward-catalog,MOST-catalog} make the SNR a member of 
TXC class. 
In the direction to MSH~11-61A lies the \g-ray source 
2EG~1103-6106 (3EG~J1102-6103) \cite{Sturner-Dermer-M96,Romero-B-T-99}. 
Is it possible to consider observed flux of the EGRET 
source directed toward the \msh as \pio decay \g-ray emission?   

The \g-flux from the source 2EG~1103-6106 in the EGRET band 
$\varepsilon_\gamma=30\div2\cdot10^4\MeV$ 
is approximated as \cite{Merck-Bertsch96}
$$
S_\gamma=(1.1\pm 0.2)\cdot10^{-9}\left({\varepsilon_\gamma\over 213\MeV}
\right)^{-2.3\pm0.2}\quad {{\rm photon}\over {\rm cm}^2\cdot {\rm s}
\cdot {\rm MeV}}\ .
$$
Thus, the  
luminosity of the source 
in $\varepsilon_\gamma\geq 100\MeV$ band is respectively  
$$
L_{\gamma,{\rm obs}}^{\geq 100}=4\cdot10^{34}\ d_{\rm kpc}^2\quad {\rm erg/s}.
$$

Most recent study on \msh \cite{Slane-et-al-msh} 
gives the parameters of the object 
presented in Table~\ref{tabl-1} (different for different distance assumed). 
There are also values of 
$\xi$ 
in the table 
which allow to adjust the expected luminosity of \msh in \pio decay 
\g-rays with
observed flux from the source 2EG~1103-6106. 
We see that moderate number density $\sim 150\ {\rm cm^{-3}}$ of cloud 
located near the one region of the remnant is enough to explain the luminocity 
of 2EG~1103-6106, by protons accelereted on the shock front of MSH~11-61A. 
Note, if we take $\nu\simeq 0.1-0.2$ \cite{Ahar-Dr-V-94} instead of the value 
used here, 
$\nu=0.03$, the density of cloud 
needs to be only $\sim 20\div40\ {\rm cm^{-3}}$.   
It is interesting that the same consideration allows also to adjust the 
\pio decay \g-luminosity  
of the source 2EG~J0618+2234 (3EG~J0617+2238) directed toward IC~443 
with the luminosity of this SNR \cite{Hn-Pet-98}.   

\begin{table}[t]
\centering
\caption[]{Parameters of \msh \cite{Slane-et-al-msh}, luminosity 
$L_{\gamma,{\rm obs}}^{\geq 100}$ of 
2EG~1103-6106 and estimations on
the proton origin \g-luminosity of the SNR. 
Presented values of $\xi$ 
allow us to satisfy condition $L_{\gamma,{\rm obs}}=L_{\gamma}$. \\ 
	}
\begin{tabular}{ccccccccc}
\hline
\noalign{\smallskip}
$d$, && Age $t$, & $\overline{R}$, & $E_{51}$, & $\overline{n}_o$, & 
 $L_{\gamma,{\rm obs}}^{\geq 100}$, &  
 $L_{\gamma}^{\geq 100}/\xi^{3/2}$, & $\xi$, \\
$\kpc$ && $10^{4}$ yrs & $\pc$ & $10^{51}\erg$ & ${\rm cm^{-3}}$ & 
 $10^{36}\ {\rm erg/s}$ & $10^{32}\ {\rm erg/s}$ & $10^2$ \\
\noalign{\smallskip}
\hline
\noalign{\smallskip}
10&&1.3&18&1  &0.27&4&3.2&5\\
7 &&0.9&13&0.4&0.27&2&1.3&6\\
\noalign{\smallskip}
\hline
\end{tabular}
\label{tabl-1}
\end{table}

\section{Conclusions}

We may expect that nucleonic component of CRs accelerated on SNR shocks 
have to reveal itself in observations. 
Which SNRs might emit \g-rays mainly via neutral pion decays?    
Presented model for TXCs makes the members of this 
class very promising candidates for the \pio decay \g-ray sources 
since a central thermal 
X-ray peaked brightness might be evidence for density gradient and high 
density in one of the regions before SNR shock. 
CR distribution inside SNR is not uniform; 
most of CR should be confined in a 
relatively thin SNR shell. CRs are accelerated not only by the
forward shock, but also by the reverse one. Consideration of these factors 
causes an 
enhancemnent of the \pio decay \g-ray flux of at least on an order of 
magnitude. 
This allows us to adjust the theoretical flux with observed one 
in at least few cases. 
So, we have enigmatic situation. There are the class of prospective sources. 
Theory can fit the \pio decay \g-ray fluxes. Why we do not have many direct 
observations of such \g-rays in order to confirm theoretical predictions about 
CR generation by SNRs? The question remains open. 

{\small 
{\it Acknowledgements.} This work was partially supported by INTAS through 
grant 99-1065. Author is thankful to the organizers for 
support which allows him to participate in the Course. 
}


\end{document}